\documentstyle[12pt,epsf]{article}
\input{tcilatex}

\begin{document}

\thispagestyle{empty}

\begin{center}
{\Large Diffraction of X-ray pulse in crystals.}$\medskip $

{\bf V.G.Baryshevsky}
\end{center}

{\bf \vspace{1pt}}

\begin{center}
Institute of Nuclear Problems, Bobruiskaya Str.11, Minsk 220080 Belarus

Electronic address: bar@inp.minsk.by

\vspace{1pt}

\bigskip
\end{center}

Recently the investigation of the extremely short (subpicosecond) X-ray
pulses  interaction with crystals takes interest because of the development
of linac-driven X-ray Free Electron Laser, operating in the SASE mode and
X-ray Volume Free Electron Laser \cite{1, 2}.

According to the analysis \cite{3} short X-ray pulse passing through a
crystal is accompanied by the significant time delay of radiation. The $%
\delta -$pulse delay law for the Bragg diffraction is proportional to $\
\sim \left| \dfrac{J_{1}(at)}{t}\right| ^{2}$ , where $J_{1}$ - is the
Bessel function, $a$ - a coefficient will be defined below, $t$ - time.

\bigskip In the present paper the delay law dependence on\ the diffraction
asymmetry parameters is analyzed. It is shown that the use of subpicosecond
pulses allows to observe the phenomenon of the time delay of pulse in
crystal  and to investigate the delay law experimentally. It is also shown
that the pulse delay law depends on the quanta polarization.

Let us consider the pulse of electromagnetic radiation passing through the
medium with the refraction index $n(\omega )$. The wave packet group
velocity is as follows:

\begin{equation}
v_{gr}=\left( \frac{\partial \omega n(\omega )}{c\partial \omega }\right)
^{-1}=\frac{c}{n(\omega )+\omega \frac{\partial n(\omega )}{\partial \omega }%
},
\end{equation}
where $c$ - is the speed of light, $\omega $ - is the quantum frequency.

In the X-ray range ( $\sim $tens of keV) the index of refraction has the
universal form $n(\omega )=1-\dfrac{\omega _{L}^{2}}{2\omega ^{2}}$ , $\
\omega _{L}$ is the Langmour frequency. Additionally, $n-1\simeq 10^{-6}\ll 1
$. Substituting \ $n(\omega )$ to (1) one can obtain that $v_{gr}\simeq
c\left( 1-\dfrac{\omega _{L}^{2}}{\omega ^{2}}\right) $. It is clear that
the group velocity is close to the speed of light. Therefore the time of
wave packet delay in a medium is much shorter than the that needed for
passing the length equal to the target width in a vacuum.

\begin{equation}
\Delta T=\frac{1}{v_{gr}}-\frac{1}{c}\simeq \frac{l}{c}\dfrac{\omega _{L}^{2}%
}{\omega ^{2}}\ll \frac{l}{c}.
\end{equation}

To consider the pulse diffraction in a crystal one should solve Maxwell
equations that describe pulse passing through a crystal. Maxwell equations
are linear, therefore it is convenient to use Fourier transform by time and
to rewrite these equations as functions of frequency: 
\begin{equation}
\left[ -curl~curl~\vec{E}_{\vec{k}}(\vec{r},\omega )+\frac{\omega ^{2}}{c^{2}%
}\vec{E}_{\vec{k}}(\vec{r},\omega )\right] _{i}+\chi _{ij}(\vec{r},\omega
)~E_{\vec{k},j}(\vec{r},\omega )=0,
\end{equation}
where $\chi _{ij}(\vec{r},\omega )$ - is the spatially periodic tensor of
susceptibility, $i,j=1,2,3,$ repeated \ indices imply summation.

Making the Fourier transformation of these equations by coordinate variables
one can derive a set of equations matching the incident and diffracted
waves. When two strong waves are excited under diffraction (so-called
two-beam diffraction case) the following set of equations for wave
amplitudes determining  can be obtained: 
\begin{equation}
\begin{array}{c}
\left( \frac{k^{2}}{\omega ^{2}}-1-\chi _{0}\right) \vec{E}_{\vec{k}%
}^{s}-c_{s}\chi _{-\vec{\tau}}\vec{E}_{\vec{k}_{\tau }}^{s}=0 \\ 
\\ 
\left( \frac{k_{\tau }^{2}}{\omega ^{2}}-1-\chi _{0}\right) \vec{E}_{\vec{k}%
_{\tau }}^{s}-c_{s}\chi _{\vec{\tau}}\vec{E}_{\vec{k}}^{s}=0
\end{array}
\end{equation}
Here $\vec{k}$ is the wave vector of the incident wave, $\vec{k}_{\vec{\tau}%
}=\vec{k}+\vec{\tau}$, $\vec{\tau}$ is the reciprocal lattice vector; $\chi
_{0},\chi _{\vec{\tau}}$ are the Fourier components of the crystal
susceptibility: 
\begin{equation}
\chi (\vec{r})=\sum_{\vec{\tau}}\chi _{\vec{\tau}}\exp (i\vec{\tau}\vec{r})
\end{equation}
$C_{s}=\vec{e}^{~s}\vec{e}_{\vec{\tau}}^{~s}$, $\vec{e}^{~s}(\vec{e}_{\vec{%
\tau}}^{~s})$ are the unit polarization vectors of the incident and
diffracted waves, respectively.

The condition for the linear system (4) to be solvable leads to a dispersion
equation that determines the possible wave vectors $\vec{k}$ in a crystal.
It is convenient to present these wave vectors as: 
\[
\vec{k}_{\mu s}=\vec{k}+\text{\ae }_{\mu s}\vec{N},~\ae _{\mu s}=\frac{%
\omega }{c\gamma _{0}}~\varepsilon _{\mu s},
\]
where $\mu =1,2$; $\vec{N}$ is the unit vector of a normal to the entrance
crystal surface directed into a crystal , 
\begin{equation}
\varepsilon _{s}^{(1,2)}=\frac{1}{4}[(1+\beta )\chi _{0}-\beta \alpha _{B}%
]\pm \frac{1}{4}\left\{ [(1+\beta )\chi _{0}-\beta \alpha _{B}-2\chi
_{0}]^{2}+4\beta C_{S}^{2}\chi _{\vec{\tau}}\chi _{-\vec{\tau}}\right\}
^{1/2},
\end{equation}
$\alpha _{B}=(2\vec{k}\vec{\tau}+\tau ^{2})k^{-2}$ is the off-Bragg
parameter ($\alpha _{B}=0$ when the Bragg condition of diffraction is
exactly fulfilled), 
\[
\gamma _{0}=\vec{n}_{\gamma }\cdot \vec{N},~~~\vec{n}_{\gamma }=\frac{\vec{k}%
}{k},~~~\beta =\frac{\gamma _{0}}{\gamma _{1}},~~~\gamma _{1}=\vec{n}%
_{\gamma \tau }\cdot \vec{N},~~~\vec{n}_{\gamma \tau }=\frac{\vec{k}+\vec{%
\tau}}{|\vec{k}+\vec{\tau}|}
\]
The general solution of (3,4) inside a crystal is: 
\begin{equation}
\vec{E}_{\vec{k}}^{s}(\vec{r})=\sum_{\mu =1}^{2}\left[ \vec{e}^{~s}A_{\mu
}\exp (i\vec{k}_{\mu s}\vec{r})+\vec{e}_{\tau }^{~s}A_{\tau \mu }\exp (i\vec{%
k}_{\mu s\tau }\vec{r})\right] 
\end{equation}

By matching these solutions with the solutions of Maxwell equation for the
vacuum area we can find the explicit expression for $\vec{E}_{\vec{k}}^{s}(%
\vec{r})$ throughout the space. It is possible to discriminate several types
of diffraction geometries, namely, the Laue and the Bragg schemes are the
most well-known \cite{4}.\bigskip\ 

In the case of two-wave dynamical diffraction crystal can be described \ by
two effective refraction indices

\[
n_{s}^{(1,2)}=1+\varepsilon _{s}^{(1,2)}, 
\]

\begin{equation}
\varepsilon _{s}^{(1,2)}=\frac{1}{4}\left\{ \chi _{{\small 0}}(1+\beta
)-\beta \alpha \pm \sqrt{(\chi _{{\small 0}}(1-\beta )+\beta \alpha
)^{2}+4\beta C_{s}\chi _{{\small \tau }}\chi _{{\small -\tau }}}\right\} .
\end{equation}

The diffraction is significant in the narrow range near the Bragg frequency,
therefore $\chi _{0}$ and $\chi _{\tau }$ can be considered as constants
and\ the dependence on $\omega $ should be taken into account for \ $\alpha =%
\dfrac{2\pi \overrightarrow{\tau }(2\pi \overrightarrow{\tau }+2%
\overrightarrow{k})}{k^{2}}=-\dfrac{(2\pi \tau )^{2}}{k_{B}^{3}c}(\omega
-\omega _{B})$, where $k=\dfrac{\omega }{c}$; $2\pi \overrightarrow{\tau }$\
- the reciprocal lattice vector that characterizes the set of planes where
the diffraction occurs; Bragg frequency is determined by the condition $%
\alpha =0$.

From (1,8)\ one can obtain

\begin{equation}
v_{gr}^{(1,2)s}=\frac{c}{n^{(1,2)}(\omega )\pm \beta \dfrac{(2\pi \tau )^{2}%
}{4k_{B}^{2}}\dfrac{(\chi _{{\small 0}}(1-\beta )+\beta \alpha )}{\sqrt{%
(\chi _{{\small 0}}(1-\beta )+\beta \alpha )^{2}+4\beta C_{s}\chi _{{\small %
\tau }}\chi _{{\small -\tau }}}}}.
\end{equation}

\bigskip In the general case $(\chi _{0}(1-\beta )+\beta \alpha )\simeq 2%
\sqrt{\beta }\chi _{0}$, therefore the term that is added to the $%
n_{s}^{(1,2)}(\omega )$ in the denominator (9) is of the order of 1.\
Moreover, $v_{gr}$ significantly differs from $c$ for the antisymmetric
diffraction $(\left| \beta \right| \gg 1).$ It should be noted that because
of the complicated character of the wave field in a crystal one of the $%
v_{gr}^{(i)s}$ can appear to be much higher than $c$\ and negative. When $%
\beta $ is negative the subradical expression in (9) can become equal to
zero (Bragg reflection threshold) and $v_{gr}\rightarrow 0$\ . It should be
noted that in the presence of the time-alternative external field a crystal
can be described by the effective indices of refraction that depend on the
external field frequency $\Omega $\ . Therefore, in this case $v_{gr}$
appears to be the function of  $\Omega $\ . This can be easily observed in
the conditions of X-ray-acoustic resonance. The analysis done allows to
conclude that center of the X-ray pulse can \ undergo the significant delay
in a crystal $\Delta T\gg \dfrac{l}{c}$ that it is possible to investigate
experimentally. Thus, when $\beta =10^{3}$, $l=0,1$ cm and $l/c\simeq 3\cdot
10^{-12}$ the delay \ time can be estimated as $\Delta T\simeq 3\cdot 10^{-9}
$sec.

\bigskip

Let us study now the time dependence of delay law of radiation \ after
passing through a crystal. Assuming that $B(\omega )$ is the reflection or
transmission amplitude coefficients of a crystal one can obtain the
following expression for the pulse form

\begin{equation}
E(t)=\frac{1}{2\pi }\int B(\omega )E_{0}(\omega )e^{-i\omega t}d\omega =\int
B(t-t^{\prime })E_{0}(t^{\prime })dt^{\prime }.
\end{equation}
where $E_{0}(\omega )$ is the amplitude of the electromagnetic wave incident
on a crystal\ \ \ \ 

\bigskip In accordance with the general theory for the Bragg geometry the
amplitude of the diffractionally reflected wave for the crystal width that
is much greater than the absorbtion length can be written \cite{4}

\begin{equation}
B_{s}(\omega )=-\frac{1}{2\chi _{\tau }}\left\{ \chi _{{\small 0}}(1+\left|
\beta \right| )-\left| \beta \right| \alpha -\sqrt{(\chi _{{\small 0}%
}(1-\left| \beta \right| )-\left| \beta \right| \alpha )^{2}-4\left| \beta
\right| C_{s}\chi _{{\small \tau }}\chi _{{\small -\tau }}}\right\}
\end{equation}

\bigskip In the absence of resonance scattering the parameters $\chi _{0}$
and $\chi _{\pm \tau }$ can be considered as \ constants and frequency
dependence is defined by the term $\alpha =-\dfrac{(2\pi \tau )^{2}}{%
k_{B}^{3}c}(\omega -\omega _{B})$. So, $B_{s}(t)$\ \ can be find from

\begin{equation}
B_{s}(t)=-\frac{1}{4\pi \chi _{\tau }}\int \left\{ \chi _{{\small 0}%
}(1+\left| \beta \right| )-\left| \beta \right| \alpha -\sqrt{(\chi _{%
{\small 0}}(1-\left| \beta \right| )-\left| \beta \right| \alpha
)^{2}-4\left| \beta \right| C_{s}\chi _{{\small \tau }}\chi _{{\small -\tau }%
}}\right\} e^{-i\omega t}d\omega .
\end{equation}

Fourier transform of the first term results in $\delta (t)$ and we can
neglet it, because the delay is described by the second term. The second
term can be calculated by the methods of theory of \ function of complex
argument:

\begin{equation}
B_{s}(t)=-\frac{i}{4\chi _{\tau }}\left| \beta \right| \dfrac{(2\pi \tau
)^{2}}{k_{B}^{2}\omega _{B}}\frac{J_{1}(a_{s}t)}{t}e^{-i(\omega _{B}+\Delta
\omega _{B})t}\theta (t),
\end{equation}

\bigskip or

\begin{equation}
B_{s}(t)=-\frac{i\sqrt{\left| \beta \right| }}{2}\frac{J_{1}(a_{s}t)}{a_{s}t}%
e^{-i(\omega _{B}+\Delta \omega _{B})t}\theta (t),
\end{equation}

where

\[
a_{s}=\frac{2\sqrt{C_{s}\chi _{\tau }\chi _{-\tau }}\omega _{B}}{\sqrt{%
\left| \beta \right| }\dfrac{(2\pi \tau )^{2}}{k_{B}^{2}}},\Delta \omega
_{B}=-\frac{\chi _{{\small 0}}(1+\left| \beta \right| )\omega _{B}k_{B}^{2}}{%
\left| \beta \right| (2\pi \tau )^{2}}. 
\]

\bigskip

Since $\chi _{0}$ and $\chi _{\tau }$ are complex, both $a_{s}$ and $\Delta
\omega _{B}$ have real and imaginary parts. According to (12-14) in the case
of Bragg reflection of short pulse (pulse frequency band  width $\gg $
frequency width of the total reflection range) appear both the instantly
reflected pulse and the pulse with amplitude undergoing damping beatings.
Beatings period increases with $\left| \beta \right| $ grows and $\chi
_{\tau }$\ decrease. Pulse intensity can be written as

\begin{equation}
I_{s}(t)\sim \left| B_{s}(t)\right| ^{2}=\frac{\left| \beta \right| }{2}%
\left| \frac{J_{1}(a_{s}t)}{at}\right| ^{2}e^{-2\func{Im}\Delta \omega
_{B}t}\theta (t).
\end{equation}

It is evident that the reflected pulse intensity depends on the orientation
of photon polarization vector $\overrightarrow{e_{s}}$ and undergoes the
damping oscillations on time. \qquad \qquad\ \ 

Let us evaluate the effect. Characteristic values are $\func{Im}\Delta
\omega _{B}\sim \func{Im}\chi _{0}\omega _{B}$ and $\func{Im}a\sim \dfrac{%
\func{Im}\chi _{\tau }\omega _{B}}{\sqrt{\beta }}.$ For 10 keV for the
crystal of Si $\ \func{Im}\chi _{0}=1,6\cdot 10^{-7}$ , \ for LiH $\ \func{Im%
}\chi _{0}=7,6\cdot 10^{-11},\func{Im}\chi _{\tau }=7\cdot 10^{-11}$, \ for
LiF $\ \func{Im}\chi _{0}\sim 10^{-8}.$ Consequently, the characteristic
time $\tau $\ for the exponent decay in (15) can be estimated as follows ($%
\omega _{B}=10^{19}$):

for Si - $\tau \sim 10^{-12}$ sec, for LiF - $\tau \sim 10^{-10}$ sec, for
LiH - $\tau \sim 10^{-9}$ sec!!

The reflected pulse also undergoes oscillations period of which increases
with $\left| \beta \right| $ grows and decreasing of $\func{Re}\chi _{\tau
}. $ This period can be estimated for $\beta =10^{2}$ and $\func{Re}\chi
_{\tau }\sim 10^{-6}$ as $T\symbol{126}10^{-12}$ sec (for Si, LiH, LiF).

\bigskip When the resolving time of the detection equipment is greater than
the oscillation period the expression (15) should be averaged over the
period of oscillations. Then, for the time intervals when $\func{Re}%
a_{s}t\gg 1,$ $\func{Im}\Delta \omega _{B}t\ll 1$ the delay law (15) has the
power function form:

\[
{\large I}_{s}{\large (t)\,\sim \,t}^{-3}{\large .}
\]


\bigskip

\bigskip

\begin{thebibliography}{9}
\bibitem{1}  V.G.Baryshevsky, K.G.Batrakov, I.Ya.Dubovskaya J.Phys. D: Appl.
Phys. 24(1991) 1250-1257.

\bibitem{2}  CERN COURIER 39, N4 (1999) \ 11-12

\bibitem{3}  V.G.Baryshevsky Izvestia AN BSSR ser.phys.-mat. N5 (1985)
109-112

\bibitem{4}  Z.G.Pinsker Dynamical scattering of X-rays in crystals
(Springer, Berlin, 1988) 
\end{thebibliography}
\end{document}